\documentclass[a4paper,12pt]{article}
\newif\ifpdf            
\ifx\pdfoutput\undefined \pdffalse \else \pdftrue \fi
\ifpdf 
\usepackage[pdftex]{graphicx}
\pdfcompresslevel9 \DeclareGraphicsExtensions{.jpg,.pdf,.png,.mps}
\else 
\usepackage[dvips]{graphicx}
\DeclareGraphicsExtensions{.eps,.ps} \fi
\usepackage{bm}

\newcommand{\beq}{\begin{equation}}
\newcommand{\eeq}{\end{equation}}
\def\matr#1{\underline{\underline{{\bm{#1}}}}}
\def\lz{\ell_{\parallel}}
\def\lp{\ell_{\bot}}

\begin{document}

\title{Spontaneous thermal expansion of nematic elastomers }

\author{A.R. Tajbakhsh and
E.M. Terentjev\footnote{Electronic mail: {\it emt1000@cam.ac.uk}} \\
Cavendish Laboratory, University of Cambridge \\ Madingley Road,
Cambridge CB3 0HE, U.K. }

\date{\today}
\maketitle

\begin{abstract}
\noindent We study the monodomain (single-crystal) nematic
elastomer materials, all side-chain siloxane polymers with the
same mesogenic groups and crosslinking density, but differing in
the type of crosslinking. Increasing the proportion of long
di-functional segments of main-chain nematic polymer, acting as
network crosslinking, results in dramatic changes in the uniaxial
equilibrium thermal expansion on cooling from isotropic phase. At
higher concentration of main chains their behaviour dominates the
elastomer properties. At low concentration of main-chain material,
we detect two distinct transitions at different temperatures, one
attributed to the main-chain, the other to the side-chain
component. The effective uniaxial anisotropy of nematic rubber,
$r(T)=\lz/\lp$ proportional to the effective nematic order
parameter $Q(T)$, is given by an average of the two components and
thus reflects the two-transition nature of thermal expansion. The
experimental data is compared with the theoretical model of ideal
nematic elastomers; applications in high-amplitude thermal
actuators are discussed in the end.
\end{abstract}
 \vspace{0.2cm}
 {61.30.-v} \ {Liquid crystals} \\
      {46.25.Hf} \ {Thermoelasticity}

\newpage

\section{Introduction}

Liquid crystalline elastomers (LCE) have attracted a significant
experimental and theoretical interest. Recent review articles
summarise much of the recent-years research
\cite{fink00,barc00,wt00,bf00,emt00}. The novel behavior of these
materials arises from a coupling between the liquid crystalline
ordering of mesogenic moieties and the elastic properties of the
polymer network. Several unusual physical effects have been
discovered in LCE: (a) Spontaneous, reversible shape changes on
heating or illumination. (b) ``Soft elasticity'' -- mechanical
deformation without (or with very low) stress. (c) Mechanical
instabilities and discontinuous stress-strain strain relations on
switching of nematic director by mechanical fields
\cite{mitch,kundler}. (d) Electrical switching of optical
properties with accompanying mechanical strains in
\cite{rbm1,rbm2}. (e) Solid phase nematohydrodynamics and unusual
rheology and dynamics, leading to anomalous dissipation,
mechanical damping and dynamic softness \cite{prl,jap}.

Theoretical work has been able to develop simple molecular models
of ideal nematic network and qualitatively describe or predict
nearly all observed phenomena with a description having no free
parameters, based only on independently measurable quantities
\cite{wt00}. This theory of an ideal nematic rubber is encompassed
in the expression for the free energy density of a nematic rubber
in response to a arbitrary general deformation expressed by the
strain tensor $\matr{\lambda}$:
\begin{equation}
F_{\rm elast}= \textstyle{\frac{1}{2}}\mu \ {\rm Tr} \left[
\matr{\ell}_0 \cdot \matr{\lambda}^{\rm T}\cdot \matr{\ell}^{-1}
\cdot \matr{\lambda} \right] , \label{Fo}
\end{equation}
where $\mu \approx n_{\rm x} k_B T$ is the rubber modulus
proportional to the crosslinking density of the network. Within
the phantom Gaussian chain approximation, the matrix $\matr{\ell}$
(as well as $\matr{\ell}_0$) represents the anisotropic
distribution of nematic chain segments after the affine
deformation $\matr{\lambda}$ is applied (or the initial anisotropy
before deformation, in the case of $\matr{\ell}_0$). The effective
chain anisotropy is the crucial factor in the description. The
matrix $\matr{\ell}$ characterises the local nematic director
$\bm{n}$ as its principal eigenvector, as well as the degree of
local order expressed by the difference between its principal
eigenvalues: in the component form $\ell_{ij} = \lp \, \delta_{ij}
+ [\lz-\lp] n_i n_j$. Accordingly, the nematic polymer chain has
the anisotropic radius of gyration, $\overline{R}_\| =
(\frac{1}{3}\lz L)^{1/2}$ and $\overline{R}_\bot = (\frac{1}{3}\lp
L)^{1/2}$ for a Gaussian chain of contour length $L$. In the
isotropic phase both principal values become equal to the single
parameter, $b$, of chain persistence length: $\ell_{ij}^{\rm iso}
= b \, \delta_{ij}$ and $\overline{R} = (\frac{1}{3} b L)^{1/2}$
along each of the three coordinate axes.

A more recent development allowed to combine the concepts of
reptation theory of entangled networks with the anisotropic nature
of nematic polymer strands \cite{Kutter2}, obtaining the ``tube
model'' corrections to the ideal nematic rubber elastic free
energy
\begin{eqnarray}
    F_{\rm elast}&=&\frac{2}{3}\mu\ \frac{2M+1}{3M+1}
\mathrm{Tr}(\matr{\ell}_0 \cdot \matr{\lambda}^{\rm T} \cdot
\matr{\ell}^{-1} \cdot \matr{\lambda})\nonumber\\
    &&+\frac{3}{2}\mu(M-1)\frac{2M+1}{3M+1}
(\langle|\matr{\ell}^{-1/2} \cdot \matr{\lambda} \cdot
\matr{\ell}_0^{1/2} |\rangle)^2
    \nonumber\\
&&+\mu(M-1)\langle\ln|\matr{\ell}^{-1/2} \cdot \matr{\lambda}
\cdot \matr{\ell}_0^{1/2}|\rangle,
    \label{Ftube}
\end{eqnarray}
where $M$ is the average number of entanglements per network
strand ($M=1$ for an ideal unentangled nematic rubber) and the
notation $\langle ... \rangle$ stands for an angular averaging of
a matrix converted with an arbitrary unit vector \cite{Kutter2}.
In a real elastomer $M$ is expected to be rather high since the
reptation diffusion and the resulting disentanglement are
restricted in a crosslinked network, unlike in the case of
uncrosslinked melt where one needs a certain chain length ($N_e$)
to form a topological knot \cite{tube}. Clearly, at high $M$
entanglements dominate the rubber modulus, as is indicated by the
second and third lines in the Eq.~(\ref{Ftube}). However, it turns
out that, although the modified elastic free energy appears
different, many key parameters and predictions of the ideal theory
remain in force.

One such equilibrium effect stands out in the properties of liquid
crystalline elastomers, being especially pronounced when a
monodomain (single crystal, permanently aligned \cite{kupfer})
nematic network is prepared. It is the anomalous thermal expansion
-- the spontaneous elongation of the material along the director
axis on cooling from the isotropic phase
\cite{schatzle89,fink-thermo}. It is rather straightforward to
derive the equilibrium uniaxial extensional strain by minimisation
of the Eq.(\ref{Fo}) where the chain anisotropy changes from
$\lz^{(0)}=\lp^{(0)}=b$, in the initial state, to a set of values
$\lz \neq \lp$ with $\lz$ along the nematic director $\bm{n}$,
which is also the principal axis of spontaneous extension (see
\cite{wt00} for detail). The result for the uniaxial expansion,
\begin{equation}
\lambda_{\rm th} = (\lz/\lp)^{1/3} ,  \label{Lth}
\end{equation}
is a sensitive function of temperature. Remarkably, the same
expression for $\lambda_{\rm th}$ is obtained from the modified
reptation theory of nematic elastomers (\ref{Ftube}). The reason
for this spontaneous uniaxial thermal expansion is the direct
coupling between the average chain anisotropy and the nematic
order parameter. Depending on this coupling, the magnitude of
$\lambda_{\rm th}$ could differ greatly (and even change sense
from extension to contraction in a polymer with oblate backbone
conformation), reaching a maximum in a main-chain nematic polymer
systems which could expand, spontaneously and reversibly, by a
factor of $\sim 3-4$ \cite{bergemann}.

In this paper we focus on the role of the single model parameter
in the above molecular theories -- the effective chain anisotropy
$\matr{\ell}$. On the one hand, the strength of the theory is that
the effective anisotropy parameter $r=\lz/\lp$ could be directly
measured by examining the thermal expansion of a uniaxial nematic
rubber, cf. Eq.(\ref{Lth}), thus leaving no free parameters and
allowing quantitative predictions. On the other hand, one
appreciates that the value of $r$ measured in a macroscopic
experiment, such as the mechanical deformation, is an effective
average parameter of the whole network, not necessarily directly
related to the persistent length anisotropy of a particular
polymer forming the network.

We examine a sequence of materials, having essentially the same
chemical structure and composition of side chain polysiloxane
nematic polymer strands but different in the type of permanent
chemical crosslinking. Maintaining the constant degree of
crosslinking, at 10\% (by reacting bonds), we continuously alter
the proportion between the two crosslinking agents: (i) flexible
di-alkeneoxybenzene units, which are thought to have only a minor
effect on the anisotropy of nematic network strands, and (ii) long
di-functional chains of main-chain nematic polymer, which create
an additional (and very high) anisotropy in the composite
material. In all cases we prepare uniformly aligned monodomain
nematic networks -- single-crystal liquid crystal elastomers
(SCLCE) in terminology of \cite{kupfer}. By measuring the
spontaneous uniaxial thermal expansion $\lambda_{\rm th}$, we
obtain the temperature-dependent effective network anisotropy
$r(T)$. A parallel thermal study (DSC) identifies the points of of
nematic phase transitions for all materials and correlates these
with the thermo-mechanical response.

We also correlate the thermal expansion (and the effective
mechanical anisotropy $r$) with the local nematic order parameter
$Q(T)$, measured by X-ray diffraction. It is obvious that there
has to be a direct relationship, with the limit $r \rightarrow 1$
at $Q \rightarrow 0$. However, the theoretical derivation of such
a correspondence is difficult because it is based on the
single-chain analysis and is strongly model-dependent, varying
from a trivial $r =(1+2Q)/(1-Q)$ for a freely-jointed chain model
to a highly non-linear $r\sim \exp \left[3/(1-Q) \right]$ in the
hairpin regime of main-chain homopolymers \cite{mainchain}. The
results show that a linear relationship between $r(T)$ and $Q(T)$
holds over a broad range of temperatures.

Finally, we conclude by discussing the possible application of
spontaneous deformation of monodomain nematic rubber in
thermo-mechanical actuators. Section~4 presents a study of a
selected optimised sample, providing the data for deformation
under load (the artificial muscle regime \cite{degen}) and for the
stress-temperature relation at fixed geometry (the dynamic
actuator regime). There are competing factors, for instance, the
amplitude and the steepness of the response are in the opposite
relation to the response rate. Comparing the materials that span
the wide range of possibilities, we propose that highly
anisotropic main-chain containing nematic elastomers are most
suitable for large-amplitude actuation when the time is a less
relevant factor, while the low-anisotropy nematic materials
provide the low-amplitude but high-frequency response suitable,
for instance, for acoustic applications.

\section{Experimental}

\begin{figure}
  \centering
\resizebox{0.8\textwidth}{!}{\includegraphics{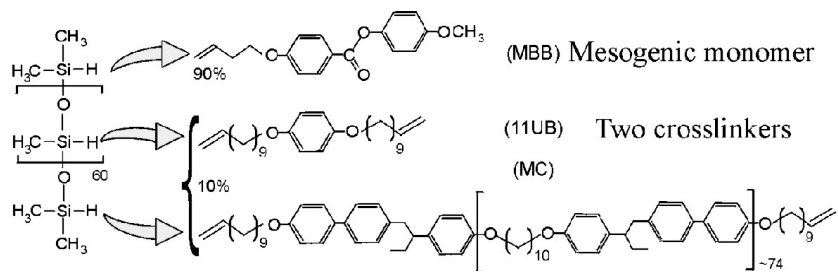}}
\caption[]{Schematic illustration of the materials used in this
work. Siloxane backbone chain with Si-H groups reacting with
90~mol\% mesogenic phenyl-benzoate side groups (MMB) and 10~mol\%
of di-vinyl crosslinking groups combining a changing proportion of
flexible small-molecule 1,4 alkeneoxybenzene (11UB) and the
main-chain nematic polymer of 1-biphenyl-2-phenyl butane (MC). }
\label{chem}
\end{figure}

All starting materials and samples of side chain siloxane liquid
crystalline elastomers were prepared in the Cavendish Laboratory
by modification of the procedure of Finkelmann et al
\cite{kupfer,greve}. The polymer backbone was a
poly-methylhydrosiloxane with approximately 60 Si-H units per
chain, obtained from ACROS Chemicals. The pendant mesogenic groups
were 4'-methoxyphenyl-4-(1-buteneoxy) benzoate (MBB), as
illustrated in Figure~\ref{chem}, attached to the backbone via the
hydrosilation reaction. All networks were chemically crosslinked
via the same reaction, in the presence of commercial platinum
catalyst COD, obtained from Wacker Chemie, with di-functional
crosslinking groups also shown in Fig.~\ref{chem} (both
synthesized in the house). In all cases the crosslinking density
was calculated to be 10~mol\% of the reacting bonds in the
siloxane backbone, so that on average each chain has 9 mesogenic
groups between crosslinking sites.

However, the samples were different in the proportion of two
crosslinking components. The first crosslinking agent, 1,4
di(11-undecene) benzene (11UB), is a small flexible molecule
deemed to have relatively minor effect on the overall mesogenic
properties of the liquid crystalline polymer (apart from the
trivial effect of added impurity, which is deliberately kept
constant in our experiments). The second crosslinking agent was
very different: di-vinyl terminated chains
$\alpha$-\{4-[1-(4'-\{11-undecenyloxy\} biphenyl)-2-phenyl]
butyl)-$\omega$-(11-undecenyloxy) poly-[1-(4-oxydeca
methyleneoxy)- biphenyl-2-phenyl] butyl chains that themselves
make a main-chain liquid crystal \cite{percec}. In our case the
crosslinking chains were with $\sim$74 rod-like monomers between
the terminal vinyl groups (determined by GPC, polydispersity
$\sim$2). Each such long-chain crosslinker connects two siloxane
polymer chains.

Calculating the crosslinking density by reacting bonds, as
described above, can be quite different from the actual
concentration of the species in resulting material. In particular,
in the case of MC polymer crosslinker, the relative gram weight of
it in the otherwise side-chain polymer matrix is very high: the
molecular weight of an average MC crosslinking chain, cf.
Fig.~\ref{chem}, is $34530 \sim 75 \times 456 $ -- which is much
greater than the molecular weight of the corresponding 9-unit
side-chain polymer strand between crosslinking points ($2691 \sim
9 \times 299 $). One may equally regard such a system as a {\em
main-chain nematic rubber network}, end-linked with relatively
small crosslinking groups made of side-chain nematic polysiloxane.
\begin{table}
\begin{center}
\begin{tabular}{|l|c|c|c|c|}
\hline
Samples  & \%(11UB) & \%(MC) &  $T_g$~(C) &  $T_{\rm ni}$~(C) \\
\hline
MC0 & 10 & 0 &  3 &  85.5 \\
MC0.1 & 9.9 & 0.1 & 7  & 85.6(106) \\
MC0.25 & 9.75 & 0.25 & 2  & 88(106) \\
MC0.5 & 9.5 & 0.5 & 7 & (91)106 \\
MC1 & 9 & 1 & 2  & (98.5)107 \\
MC10 & 0 & 10 & 17  & 108 \\
\hline
\end{tabular}
\caption{Proportions of crosslinkers 11UB and MC in the overall
crosslinking composition (of 10~mol\%) and the corresponding
temperatures of glass and nematic-isotropic transitions. Values of
$T_{\rm ni}$ in brackets refer to the second transition observed
in ``mixed'' materials. The glass transition temperature are
approximate, with an error of at least $\pm 5^{\rm o}$.}
\label{tab1}
\end{center}
\end{table}

The series of materials, synthesized in this way, are referred to
by the proportion of MC crosslinker in the total crosslinking
composition. In this way, the sample labeled as MC0 is crosslinked
purely by the 10~mol\% of small-molecule group 11UB. At the
opposite end, MC10 refers to the material crosslinked purely by
the long main-chain nematic polymer. We find particularly
interesting effects occurring at very small proportion of
main-chain crosslinker. The borderline case, when the overall mass
of MC material in the network composition is approximately equal
to that of side-chain polysiloxane, is at approximately 1.75 mol\%
of MC. We find that compositions close to this are optimal in
balancing the increasing amplitude of thermal expansion and the
decreasing rate of response. The samples we have studied, MC0.1,
MC0.25, MC0.5 and MC1, refer to the crosslinking composition
listed in the Table~1, together with their characteristic
transition temperatures. For comparison, the pure uncrosslinked
melt of the same main-chain polymer, studied in some detail in
\cite{florence}, has $T_g \approx $38~C and $T_{\rm ni} \approx$
112~C (for the longer-chain MC component).

Monodomain, aligned samples were made by preparing partially
crosslinked films in a centrifuge, highly swollen in toluene
(2-3~ml per 1~g of material), reacting for 25-35 minutes before
evaporating the solvent and suspending the samples under load in
an oven to complete the reaction for more than 5 hours at 120~C. A
careful study of reaction kinetics ensured that approximately 50\%
of crosslinks were established in the first stage of this
preparation. When a uniaxial stress is applied to such a partially
crosslinked network, the aligned monodomain state in the resulting
nematic elastomer is established with the director along the
stress axis. This orientation is then fixed by the subsequent
second-stage reaction, when the remaining crosslinks are fully
established. Following the ideas and results of \cite{kupfer}, in
all cases we performed the second-stage crosslinking in the
high-temperature isotropic phase: in this way a better alignment
and mechanical softness are achieved (in contrast to the
crosslinking in a stretched polydomain nematic phase
\cite{kupfer,frid}, which results in a number of defects and
domain walls frozen in the material).

Equilibrium transition temperatures given in the Table~1 were
determined on a Perkin Elmer, Pyris 1 differential scanning
calorimeter (DSC), extrapolating to low cooling rates, and the
nematic phase identified by polarizing optical microscopy and
X-ray scattering. Thermal expansion measurements were made by
suspending the samples, without load, in a glass-fronted oven and
measuring the variation in natural length of the samples with
temperature, $L(T)$. Precise length measurements were made using a
traveling microscope, as the samples were slowly cooled (at a rate
0.5$^{\rm o}$/min) after the complete annealing in the isotropic
state.

The force measurements were performed on a custom built device
consisting of a temperature compensated stress gauge and
controller(UF1 and AD20 from Pioden Controls Ltd) in a
thermostatically controlled chamber (Cal 3200 from Cal Controls
Ltd). The rectangular shaped samples (approximately $30\times
5\times 0.3$~mm) were mounted with clamps, at room temperature, in
such a way that the sample length $L$ remains fixed throughout the
experiment. The samples were then heated (at a rate $\sim 0.5^{\rm
o}$/min) to the isotropic phase. The stress and temperature values
were acquired by connection to a Keithley multimeter (2000 series)
and stored on a PC over an IEEE interface. The increasing force
was measured by the stress gauge and later correlated with the
effective strain in the sample, due to the decreasing underlying
natural length $L(T)$. Since the effective strain reached high
values, $\lambda_{\rm th} \sim 3-3.5$ when $T_{\rm ni}$ was
reached, many samples broke or tore. However, a good
reproducibility of force-temperature, as well as stress-strain
relations allowed one to extrapolate the data over the whole
temperature range.

\section{Thermal and mechanical equilibrium}

\subsection*{Phase transitions}

\begin{figure}
  \centering
\resizebox{0.85\textwidth}{!}{\includegraphics{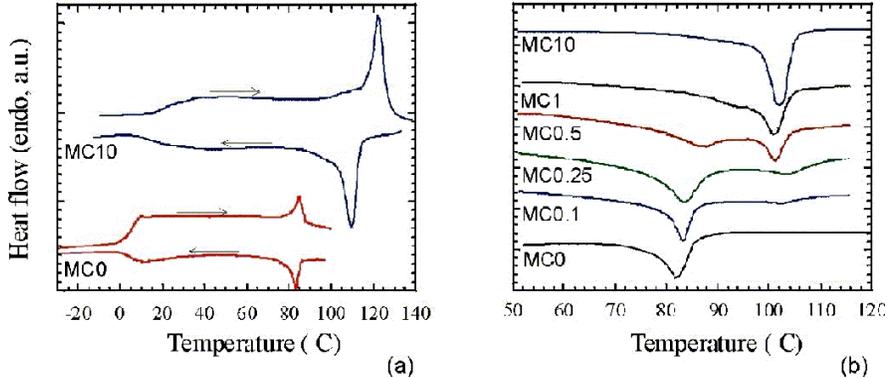}}
\caption[]{DSC data for the series of nematic elastomers, at
10~C/min. (a) Heating and cooling scans of MC0 and MC10, showing
the glass and the nematic-isotropic transitions and their
hysteresis. (b) Sequence of cooling scans of the nematic
transition for all the materials in the series, MC0 to MC10 --
increasing curves.} \label{dsc}
\end{figure}

Figure~\ref{dsc} presents the data of DSC measurements of phase
transformations in our materials. All samples exhibit a single
nematic phase above their glass transition, however, the position
and the morphology of phase transformations strongly depends on
the composition. It is expected, and well-known, that liquid
crystalline polymers show a much broader and diffuse signature of
the weakly first-order nematic-isotropic transition. This effect
is especially pronounced in crosslinked elastomers, where the
quenched sites of random disorder introduced by crosslinks lead to
a more diffuse transition and stronger hysteresis.
Fig.~\ref{dsc}(a) shows an example of heating and cooling scans
for a selected sample, MC0 and MC10, at the opposite ends of the
series. One can clearly recognize the effect of main-chain polymer
component, which increases the glass and the nematic-isotropic
transition temperatures.

Figure~\ref{dsc}(b) focuses only on the region of nematic
transition and brings together the DSC scans for all materials,
obtained at the same cooling rate of 10~C/min after a long
annealing at 120~C. Comparison between MC10 and MC1 tells that
even at the relative proportion of MC crosslinking chains of 1\%
the properties of the resulting network are dominated by their
effect. This observation will be supported below by the comparison
of spontaneous thermal expansion of MC10 and MC1, which has the
same overall character and the transition point, but rather
different from the other members of the series. One should expect
this, since the calculation of the relative mass of the side- and
the main-chain components in our materials becomes approximately
equal at the level of 1.75~mol\% of MC (see Table~1).

The second key result of this DSC study, evident in all materials
between MC0.1 and MC1, is the presence of two sequential nematic
transitions. We have verified by X-ray and optical observations
that no additional phase exists in the materials, which are
optically transparent and homogeneous at all temperatures.
Clearly, the higher-temperature transition is related to the
main-chain nematic component of the network: its temperature
hardly shifts on composition (remember that the data in
Fig.~\ref{dsc}(b) is obtained on cooling after the complete
annealing), but its overall latent heat decreases in proportion
with the decreasing percentage of MC. The second,
lower-temperature transition is, of course, that of the side-chain
polysiloxane polymer. Its temperature continuously shifts up on
increasing proportion of the MC crosslinker and, accordingly, the
stronger nematic field provided by the more strongly anisotropic
MC polymer (which has already undergone the phase transformation,
as indicated by thermal scans). We shall now investigate the
related mechanical response and return to the discussion of the
possible reasons for such a behavior.

\subsection*{Uniaxial thermal expansion}

\begin{figure}
  \centering
\resizebox{0.85\textwidth}{!}{\includegraphics{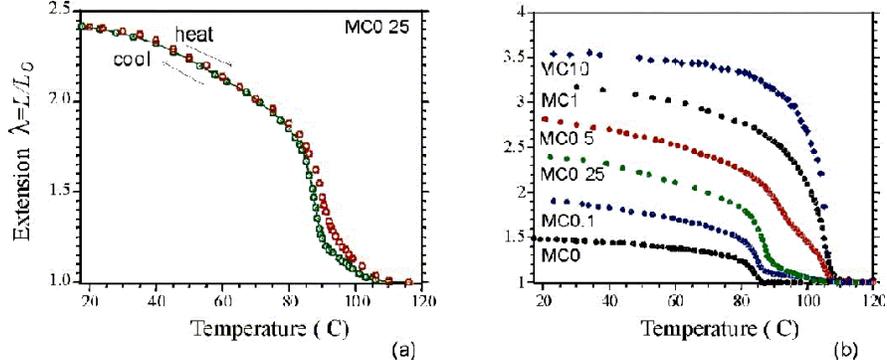}}
\caption[]{Uniaxial thermal expansion $\lambda_{\rm th}=L(T)/L_0$
for the uniaxial nematic rubber, with the temperature rate of
$\sim 0.5$~C/min. (a) Heating and cooling cycles for MC0.25,
showing the effect of hysteresis. (b) Sequence of cooling cycles
for all materials, from MC0 to MC10. Notice the two-stage
expansion, beginning at the higher nematic transition temperature
and enhanced at the lower $T_{\rm ni}$ -- consistent with the
two-transition feature of DSC data. } \label{thermal}
\end{figure}

Changing the temperature affects the nematic order parameter
$Q(T)$ of a nematic elastomer and, in a permanently aligned
monodomain state, alters the effective chain anisotropy of the
rubber, $r=\lz/\lp$. Figure~\ref{thermal} shows the variation of
natural length of free, unloaded samples, measured along the
nematic director. Obviously, in an incompressible rubber the
volume is conserved and the two perpendicular directions
experience the symmetric contraction by $1/\sqrt{\lambda}$ when
the principle direction is elongated by $\lambda=L/L_0$. Here the
length $L(T)$ is measured with respect to the length $L_0$ the
corresponding sample has in the isotropic phase. Elongation on
cooling into the nematic phase means that the network is
characterised by an effective prolate anisotropy, that is $r > 1$,
below $T_{\rm ni}$. The spontaneous uniaxial thermal expansion
described by the strain $\lambda_{\rm th}$, Eq.(\ref{Lth}), is an
equilibrium reversible process. Fig.~\ref{thermal}(a) illustrates
this by showing the evolution of sample length for one selected
material, MC0.25, on heating and on cooling. There is an evident
hysteresis which, however, is rather small at the cooling rate
used in this experiment. Without going into greater detail, we
must mention that the characteristic response time, or
equivalently -- the width of hysteresis, increase dramatically
with the increasing MC concentration. In the material MC0 we find
no trace of a hysteresis, while in MC10 the response to changing
temperature is rather slow and the hysteresis in $L(T)$ is very
noticeable at an already slow cooling/heating rate of $0.5^{\rm
o}$/min. This slow response and relaxation of main-chain nematic
polymers \cite{florence}, the likely reason being the freezing of
chain motion in narrowly confining hairpins.

Analyzing the sequence of $\lambda_{\rm th}(T)$ data for the full
series of samples, one is impressed by the high value of strain
achieved, especially in the MC-containing materials. The arguments
already presented above, about the relative mass of the long-chain
MC material in the overall network composition becoming
approximately equal to that of the siloxane side-chain polymer at
the level of $\sim$MC-1.75\%, are confirmed by the observation
that no significant further increase in the effective network
anisotropy $r=\lambda_{\rm th}^3$ occurs in high-MC containing
samples; the increase in $\lambda_{\rm th}$ at a given $T$ in
going from MC1 to MC10 is much less than from MC0 to MC1. One
could argue that the ambient-temperature value of $r\approx 42$
found in MC10 is close to the natural anisotropy of our MC
polymer. This would make a prediction for the radius of gyration
anisotropy $R_\|/R_\bot \approx 6.5$ (since $\langle R^2 \rangle
\sim \ell\, L$), which is a high but not unexpected value for this
type of semiflexible polymer, as seen by neutron scattering from
the melt (see, e.g., \cite{dallest} and references therein).

Another remarkable aspect of the results shown in the
Fig.~\ref{thermal}(b) is the dual-transition feature of the
transition, similar to that seen in the DSC study. Networks with
the low MC-concentration, when the main-chain component is the
minority in the composition, show their first sign of nematic
transition (indicated here by the initial small uniaxial
expansion) at the higher $T_{\rm ni}$, followed by a subsequent
rapid increase at the lower $T_{\rm ni}$. The amount of average
network anisotropy generated at the first transition increases
with the concentration of MC, and is consistent with the overall
latent heat of this higher-temperature transition in the
Fig.~\ref{dsc}(b). The second transition, seen as a near-critical
jump in $\lambda_{\rm th}$ for MC0.1, moves towards higher
temperatures as the MC-concentration increases, again consistently
with the DSC data.

One fundamental question that arises from the results shown above
is the nature of the first, higher $T_{\rm ni}$, nematic
transition. A natural explanation would be the phase separation of
very long MC chains during the synthesis, when they are in the
minority composition. It would then be natural that small volumes
with the high concentration of MC undergo the nematic transition
at $\sim 108$~C. The high uniaxial elongation of these small
MC-regions would, however, be absorbed by the still isotropic
network of polysiloxane chains and the overall mechanical strain
would be small. This would also explain the shift of the $T_{\rm
ni}$, because the uniaxially aligned nematic MC-regions would
provide a driving field for the rest of the mesogenic material.
However, it is not straightforward to reconcile this hypothesis
about the phase separation with the fact that our materials show
no optical inhomogeneity (which could simply mean that the
MC-regions are too small), as well as with the fact that each
MC-chain has to be crosslinked at both ends to a siloxane strand.
Since the first-step crosslinking takes place in the presence of
solvent in a well-mixed centrifuge environment, one expects a
relatively homogeneous distribution of junction points, which
should then be preserved after the second crosslinking stage, in
spite of de-swelling. An intriguing alternative possibility could
be that we observe the precursor to the nematic transition in
essentially separate chains of MC. Theoretical studies \cite{n123}
have long ago predicted the existence of nematic-nematic phase
transition lines when two competitive ordering trends are present
in the system; in that terminology, we could be observing the
discontinuous $N_{\rm III}$-$N_{\rm III}'$ transition at a
temperature below $T_{\rm ni}$ for the change between the
isotropic and the $N_{\rm III}$ phases. We have not pursued this
question in any more depth here, instead concentrating on the
mechanical aspects of nematic ordering in rubber.

\section{Thermo-mechanical actuation}

An important practical question of spontaneous thermal expansion
of monodomain nematic rubbers, relevant to practical use of this
effect in thermo-mechanical actuators, is the amount of stress the
material can sustain. The work cycle of a corresponding artificial
muscle depends on the amplitude of motion,  but also on the force
the nematic rubber exerts -- or the load it can work under. In
this section we focus on the sample MC1, which shows an almost
maximal strain $\lambda_{\rm th}(T)$, but still has a reasonably
fast response and small hysteresis.

\begin{figure}
  \centering
\resizebox{0.85\textwidth}{!}{\includegraphics{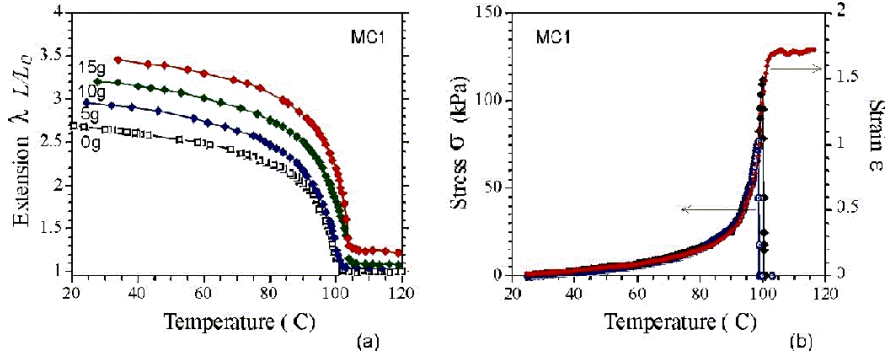}}
\caption[]{(a) Cooling cycles for MC1 under increasing load,
corresponding to applied stress of $0, 30, 60$ and $90$~kPa. (b)
The nominal stress exerted by a clamped strip of MC1, of fixed
length $L_{\rm fix}$, heated from room temperature to above
$T_{\rm ni}$. Symbols are the data points from different samples
which were tearing at certain levels of stretching, at a stress
$\sim 100$~kPa; the solid line is the plot of underlying effective
strain $\varepsilon = [L_{\rm fix}-L(T)]/L(T)$ (values on the
second $y$-axis). } \label{load}
\end{figure}

Figure~\ref{load}(a) shows the difference in thermal expansion
curves when the sample was loaded with increasing weights. For the
sample of cross-section area $\sim 1.5\,\hbox{mm}^2$ each 5~g of
weight corresponds to the stress $\sigma \approx 30$~kPa; the
curves in the plot are for the stress increasing from $0$ to
$90$~kPa. Clearly, the qualitative nature of the effect is
preserved when changing the temperature under load. An increase in
the strain values is the natural consequence of rubber elasticity:
at any given temperature and natural length, the applied load
would cause an extra stretching of the rubber band. Several
interesting effects are seen in this plot, in particular, the
small but unambiguous shift of the nematic transition temperature,
$\Delta T_{\rm ni}\sim 5^{\rm o}$, with increasing load. Such an
effect is expected on theoretical grounds in a weak first-order
transition system in an external field -- however, in ordinary
liquid crystals, electric and magnetic fields are usually too
small to create a noticeable shift. On the other hand, there is no
trace of supercritical behaviour or paranematics phase above
$T_{\rm ni}$ \cite{claudia}: the transition remains abrupt and
well-defined; the small increase of $\lambda$ above $T_{\rm ni}$
is purely due to the rubber-elastic response to the applied load.

Another practically important experiment is reported in
Fig.~\ref{load}(b). Here we measure the force that a nematic
rubber band can exert when thermally contracting. A strip of MC1
elastomer with a cross-section of $5 \times 0.3$~mm (standard in
all experiments described in this paper) is rigidly clamped in a
dynamometer device at ambient temperature, so that its length $L$
remains constant. On heating, the underlying natural length
decreases, $L_0(T)$, according to the data of the previous
section, cf. Fig.~\ref{thermal}(b) for MC1. Accordingly, the
effective extensional strain increases in the material and the
sample exerts an increasing force on its clamped ends. This force
is measured and transformed into the nominal stress by dividing by
the sample cross-section area.

The plot in Fig.~\ref{load}(b) shows two separate experiments on
heating the constrained rubber band MC1, circles and diamonds data
points (the reproducibility of results is comforting). In both
cases the sample broke before reaching the isotropic phase --
clearly, the stress of the order of 100~kPa has been the limit
this particular material could bear (one could conceivably
increase the strength by tailoring the sample shape and in
particular, its thickness). We performed a number of such
measurements and the results are very reproducible for a given
material and sample shape. The second $y$-axis on this plot gives
the variation of the underlying effective strain. We calculate it
determining the actual gradient of underlying displacement with
respect to the ``virtual'' equilibrium length, $\varepsilon =
[L_{\rm fix}-L(T)]/L(T)$, where $L_{\rm fix}$ is device-fixed,
while $L(T)$ follows the data measured and reported in the
previous section. One can clearly see that the internal stress and
the effective strain follow exactly the same graph and, therefore,
are linear functions of each other over a large region of
deformations.

\section{Conclusions}

In this article we report the systematic study of uniaxial thermal
expansion of monodomain nematic elastomers. Composite materials
combined main-chain and side-chain nematic polymers in their
networks and the results are sensitive functions of concentration
of the MC component. Increasing the proportion of MC in the
otherwise side-chain nematic network has a great effect on the
average effective anisotropy of polymer chains and, as a result,
on the magnitude of spontaneous strain $\lambda_{\rm th}$. At the
same time, there is a pronounced slowing down of all response
processes: the stress relaxation, the equilibration of natural
length and the shape recovery. One does expect such slow dynamics
in MC-dominated materials, which is consistent with other studies
of main-chain nematic polymers. The isotropic-nematic transition
at $T_{\rm ni}$ is abrupt in all our materials, with no
supercritical effects or paranematic phase. However, both the
mechanical and the thermal data indicate that at a small
concentration of MC, the network experiences a sequence of two
nematic transitions. A possible explanation for this effect could
be a microphase separation of MC-rich regions, however, there
factors against it and the question is far from being fully
understood.

The study of mechanical actuation, that is the contraction under
load, or exerting a force on changing the temperature shows that
uniformly aligned nematic elastomers have a great potential in
practical applications, from stress and deformation gauges, to
artificial muscles and micromanipulators. On the map of material
properties of different materials used in actuator design
\cite{ashby}, the nematic elastomers would occupy the region of
low stress (unlikely to exceed a few~MPa) and extremely high
deformations. In comparison, the present ``large-amplitude''
actuators are based on shape-memory alloys, which reach strains of
up to 10\% -- something that pales in comparison with over-300\%
deformations described in this work.

\subsection*{Acknowledgements}

This work has been supported by EPSRC UK. We are grateful to S.M.
Clarke, M. Warner and A. Hotta for a number of useful discussions
and help with the measurements. Special thanks to H. Finkelmann
and A. Greve for advice and help with material synthesis
procedures.


\newpage

\begin{thebibliography}{10}
\bibitem{fink00}W. Gleim and H. Finkelmann, in: {\it Side-Chain Liquid
Crystal Polymers}, ed. C.~B. McArdle (Blackie \& Sons, 1989)
p.287.
%
\bibitem{barc00}G.~G. Barclay and C.~K. Ober, {\it Progr. Polym.
Sci.} {\bf 18} 899 (1993).
%
\bibitem{wt00}
M. Warner and E.~M. Terentjev, Prog. Polym. Sci., {\bf 21}, 853
(1996).
%
\bibitem{bf00}
H.~R. Brand and H. Finkelmann, in: {\it Handbook of Liquid
Crystals}, ed D. Demus et al. (Wiley-VCH, Weinheim, 1998), Vol.3,
Chapter V.
%
\bibitem{emt00}
E.~M. Terentjev, {\it J. Phys. Cond. Mat.}, {\bf 11}, R239 (1999).
%
\bibitem{uv}H. Finkelmann, E. Nishikawa, G.~G. Pereira and M.
Warner, {\it Phys. Rev. Lett.} -- to appear
%
\bibitem{mitch}G.~R. Mitchell, F.~J. Davis and W. Guo,
 {\it Phys. Rev. Lett.} {\bf 71} 2947 (1993).
%
\bibitem{kundler}H. Finkelmann, I. Kundler, E.~M. Terentjev and M. Warner,
{\it J. Physique II}, {\bf 7}, 1059, (1997).
%
\bibitem{rbm1}C.-C. Chang, L.-C. Chien and R.~B. Meyer, {\it Phys. Rev. E}
{\bf 56}, 595 (1997).
%
\bibitem{rbm2}E.~M. Terentjev, M. Warner, R.~B. Meyer and J. Yamamoto,
{\it Phys. Rev. E} {\bf 60}, 1872, (1999).
%
\bibitem{prl}S.~M. Clarke, A.~R. Tajbakhsh, E.~M. Terentjev and M. Warner,
{\it Phys. Rev. Lett.} {\bf 86}, 4044 (2001).
%
\bibitem{jap}S.~M. Clarke, A.~R. Tajbakhsh, E.~M. Terentjev, C. Remillat,
G.~R. Tomlinson and J. House, {\it J. Appl. Phys.} {\bf 89}, 6530
(2001).
%
\bibitem{Kutter2}S. Kutter and E.~M. Terentjev, {\it Euro. Phys.
J. B} -- to appear (cond-mat/0106193).
%
\bibitem{tube}L. Fetters, D. Lohse, D. Richter, T. Witten and A.
Zirkel, {\it Macromolecules}, {\bf 27}, 4639 (1994).
%
\bibitem{kupfer}J. K\"upfer and H. Finkelmann, {\it Macromol.
Rapid Comm.} {\bf 12}, 717 (1991).
%
\bibitem{schatzle89}J. Sch\"atzle, W. Kaufhold and H. Finkelmann,
{\it Makromol. Chem.} {\bf 190}, 3269 (1989).
%
\bibitem{fink-thermo}N. Assfalg and H. Finkelmann,
{\it Kaut. Gummi. Kunstst.} {\bf 52},  677 (1999).
%
\bibitem{bergemann}G.~H.~F. Bergmann, H. Finkelmann, V. Percec, and M. Zhao,
{\it Macromol. Rapid. Comm.} {\bf 18}, 353 (1997).
%
\bibitem{mainchain}M. Warner and X.~J. Wang, Macromolecules {\bf 24}, 4932 (1991).
%
\bibitem{degen}P.~G. de Gennes, M. Hebert and R. Kant, {\it Macromol. Symp.}
{\bf 113}, 39 (1997).
%
\bibitem{greve}H. Finkelmann, A. Greve and M. Warner, {\it Euro.
Phys. J. E} {\bf 5}, 281 (2001).
%
\bibitem{percec}V. Percec, M. Kawasumi, {\it Macromolecules},
{\bf 24}, 6318 (1991).
%
\bibitem{frid}S.~V. Fridrikh and E.~M. Terentjev,
{\it Phys. Rev. E} {\bf 60}, 1847, 1999.
%
\bibitem{florence}F. Elias, S.~M. Clarke, R. Peck and E.~M. Terentjev,
{\it Europhys. Lett.} {\bf 47}, 442, (1999).
%
\bibitem{dallest}M.~H. Li, A. Brulet, P. Davidson, P. Keller,
J.~P. Cotton, {\it Phys. Rev. Lett.} {\bf 70}, 2297 (1993).
%
\bibitem{n123}W. Renz and M. Warner, {\it Proc. Roy. Soc.
Lond. A Mat} {\bf 417}, 213 (1988).
%
\bibitem{claudia}S. Disch, C. Schmidt and H. Finkelmann, {\it
Macromol. Rapid Comm.} {\bf 15}, 303 (1994).
%
\bibitem{ashby}M.~F. Ashby, {\it Materials Selection and Process in Mechanical
Design.} (Butterworth Heinemann, Oxford, 1999).

\end{thebibliography}
\end{document}